\documentclass[conference,a4paper]{IEEEtran}
\IEEEoverridecommandlockouts
\usepackage{cite}
\usepackage{amsmath,amssymb,amsfonts,amsthm}
\usepackage{algorithmic}
\usepackage{graphicx}
\usepackage{textcomp}
\usepackage{algorithm}
\usepackage{multirow}
\usepackage{caption}
\usepackage{xcolor,soul}
\usepackage{bm}
\newtheorem{remark}{Remark}
\newtheorem{definition}{Definition}
\def\BibTeX{{\rm B\kern-.05em{\sc i\kern-.025em b}\kern-.08em
    T\kern-.1667em\lower.7ex\hbox{E}\kern-.125emX}}
\newtheorem{theorem}{Theorem}
\begin{document}

\title{Computation of Rate-Distortion-Perception Functions With Wasserstein Barycenter
}


\def\@IEEEauthorblockconfadjspace{-0.25em} 
\author{\IEEEauthorblockN{Chunhui Chen\IEEEauthorrefmark{1}\IEEEauthorrefmark{2},
Xueyan Niu\IEEEauthorrefmark{2},
Wenhao Ye\IEEEauthorrefmark{1}\IEEEauthorrefmark{2},
Shitong Wu\IEEEauthorrefmark{1}\IEEEauthorrefmark{2},
Bo Bai\IEEEauthorrefmark{2},
Weichao Chen\IEEEauthorrefmark{2},
Sian-Jheng Lin\IEEEauthorrefmark{2},
}
  
\IEEEauthorblockA{\IEEEauthorrefmark{1}Department of Mathematical Sciences, Tsinghua University, Beijing, China}
\IEEEauthorblockA{\IEEEauthorrefmark{2}Theory Lab, 2012 Labs, Huawei Technologies Co., Ltd., Hong Kong, China \\
cch21@mails.tsinghua.edu.cn, niuxueyan3@huawei.com, \{yewh20, wust20\}@mails.tsinghua.edu.cn, \\ \{baibo8, chenweichao5, lin.sian.jheng1\}@huawei.com} \vspace{-2em}
}

\maketitle

\begin{abstract}
The nascent field of Rate-Distortion-Perception (RDP) theory is seeing a surge of research interest due to the application of machine learning techniques in the area of lossy compression. 
The information RDP function characterizes the three-way trade-off between description rate, average distortion, and perceptual quality measured by discrepancy between probability distributions.
%
%
However, computing RDP functions has been a challenge due to the introduction of the perceptual constraint, and existing research often resorts to data-driven methods.
In this paper, we show that the information RDP function can be transformed into a Wasserstein Barycenter problem.
%
%
%
The non-strictly convexity brought by the perceptual constraint can be regularized by an entropy regularization term.
We prove that the entropy regularized model converges to the original problem.
%
Furthermore, we propose an alternating iteration method based on the Sinkhorn algorithm to numerically solve the regularized optimization problem. 
Experimental results demonstrate the efficiency and accuracy of the proposed algorithm.
\end{abstract}


\section{Introduction}

Lossy compression plays a vital role in the communication and storage of images, videos, and audio data \cite{DBLP:conf/nips/MentzerTTA20,DBLP:journals/tcsv/MaZJZWW20,lu2020end,DBLP:journals/taslp/ZeghidourLOST22}. 
As the cornerstone of lossy compression,
the classical Rate-Distortion (RD) theory \cite{shannon1959coding} studies the tradeoff between the bit rate used for representing data and the distortion caused by compression \cite{DBLP:books/wi/01/CT2001}. 
The reconstruction quality is traditionally measured by a per-letter distortion metric, such as the mean-squared error. However, it has been shown that in practice, minimizing the distortion does not result in perceptually satisfying output for human subjects \cite{DBLP:journals/corr/SanturkarBS17}. 
%
Since high perceptual quality may come at the expense of distortion \cite{DBLP:conf/iccv/AgustssonTMTG19,DBLP:conf/cvpr/BlauM18}, researchers are motivated to extend the RD theory by bringing perception into account \cite{blau2019rethinking,zhang2021universal,niu2023conditional}.
%

Blau and Michaeli first proposed and studied the information Rate-Distortion-Perception (RDP) functions in \cite{blau2019rethinking}. Theoretical solutions with closed form expressions to the RDP problem are often intractable, except for some special cases such as the Gaussian source with squared error distortion and Wasserstein-2 metric perception \cite{zhang2021universal}. 
Therefore, a computation method for RDP functions is desirable.
%

Traditionally, the Blahut–Arimoto (BA) algorithm \cite{DBLP:journals/tit/Arimoto72,DBLP:journals/tit/Blahut72} has been successful in the computation of capacities and RD functions.
%
%
However, to our best knowledge, we have not seen any generalization of the BA algorithm to computing RDP functions. 
This may be due to the fact that RDP functions have more independent variables than RD functions, while in each alternating iteration step in the BA algorithm, all variables except the updating one need to be fixed. 
Also, RDP functions own an additional nonlinear constraint on the perception which destroys the original simplex structure and invalidates existing numerical methods. 
An alternative way to compute RDP functions is based on data-driven methods, 
%
%
such as generative adversarial networks, which minimize a weighted combination of the distortion and the perception \cite{blau2019rethinking,zhang2021universal,DBLP:conf/pcs/KirmemisT21}.
However, these deep learning-based methods often require huge computational resources while having poor generalization ability. 
%

In this paper, we propose a numerical method for computing the RDP functions. 
Referring to the approach in a recent paper \cite{wu2022communication} of RD, we reformulate RDP functions in an Optimal Transport (OT) form with an additional constraint on perception. 
However, compared to RD functions, the introduction of an additional constraint on perceptual quality destroys the origin simplex structure in RDP functions. 
%
Thus the Alternating Sinkhorn (AS) algorithm proposed in \cite{wu2022communication} cannot be applied directly to solving RDP functions. 
To handle this issue, we prove that the additional constraint can be equivalently converted to a set of linear constraints by introducing an auxiliary variable. 
Consequently, we observe that the new model appears to be in the form of the celebrated Wasserstein Barycenter problem \cite{pass2015multi,DBLP:conf/icml/CuturiD14,DBLP:journals/siamma/AguehC11}, as it can be viewed as a minimizer over two couplings (i.e., the transition mappings and the newly introduced variable) between Barycenter (i.e., the reconstruction distribution) and the source distribution. Moreover, the objective of the optimization is to compute a weighed distance according to the Wasserstein metric.
%
%
Our model therefore will be referred to as the Wasserstein Barycenter model for Rate-Distortion-Perception functions (WBM-RDP). 

With the operations above, we are able to design an algorithm for the WBM-RDP. 
In order to tackle the difficulty that the WBM-RDP is not strictly convex, we construct an entropy-regularized formulation of the WBM-RDP. We show that the new form admits a unique optimal solution, and that it converges to the origin WBM-RDP.
After obtaining the Lagrangian of the entropy regularized WBM-RDP, we observe that the degrees of freedom therein can be divided into three groups to be optimized alternatively with closed-form solutions. 
As such, we propose an improved AS algorithm, which effectively combines the advantages of AS algorithm for RD functions \cite{wu2022communication} and entropy regularized algorithm for Wasserstein Barycenter model \cite{DBLP:conf/icml/CuturiD14}. 
%
%
Numerical experiments demonstrate that the proposed algorithm reaches high precision and efficiency under various circumstances. 

\section{RDP functions and WBM-RDP}



\subsection{Rate-Distortion-Perception Functions}
Consider a discrete memoryless source $X \in \mathcal{X}$ and a reconstruction $\hat{X} \in \hat{\mathcal{X}}$, where $\mathcal{X}=\{x_1,\cdots, x_M\},\hat{\mathcal{X}}=\{\hat{x}_1,\cdots, \hat{x}_N\}$ are finite alphabets. 
Suppose $p_X$ and $p_{\hat{X}}$ are defined on the probability space $(\mathcal{X}, \mathcal{F}, \mathbb{P}).$ We consider the single-letter distortion measure
$\Delta: \mathcal{X}\times \hat{\mathcal{X}}\mapsto [0,\infty)$ and perception measure between distributions
$d:\mathbb{P}\times \mathbb{P}\mapsto [0,\infty).$

\begin{definition}[The information RDP function \cite{blau2019rethinking}]
Given a distortion fidelity $D$ and a perception quality $P$, the information RDP function is defined as\par 
%
%
%
%
\begin{small}
\begin{subequations}\label{eq0}
\begin{align}
R(D, P)= \min _{p_{\hat{X} \mid X}} \quad& I(X, \hat{X}) \label{eq0_a} \vspace{1ex} \\
 \text { s.t. }\quad &\mathbb{E}[\Delta(X, \hat{X})] \leq D,\label{eq0_b} \vspace{1ex} \\  
 &  d\left(p_X, p_{\hat{X}}\right) \leq P, \label{eq0_c}
\end{align}
\end{subequations}%
\end{small}%
where the minimization is taken over all conditional distributions, and $I(X,\hat{X})$ is the mutual information. 
%
%
\end{definition}
Note that the information RDP function \eqref{eq0} degenerates to RD functions when the constraint \eqref{eq0_c} is removed.

Since the alphabets $\mathcal{X}$ and $\hat{\mathcal{X}}$ are finite, we denote 
\[p_i = p_X(x_i),\ r_j = p_{\hat{X}}(\hat{x}_j),\  d_{ij} = \Delta(x_i,\hat{x}_j)\]
and $w_{ij}=W(\hat{x}_j \mid x_i)$ for all $1\le i\le M, 1\le j\le N$. Here $W: \mathcal{X} \rightarrow \hat{\mathcal{X}}$ is the channel transition mapping.
Thus the discrete form of problem (\ref{eq0}) can be written as
%
\begin{subequations} \label{eq0_0}
\begin{align}
    \min _{\bm{w},\bm{r}} \quad  &\sum_{i=1}^M \sum_{j=1}^N\left(w_{i j} p_i\right)\left[\log w_{i j}-\log r_j\right] \vspace{1ex} \label{eq0_0_a}\\
    \text { s.t. }  &\sum_{j=1}^N w_{i j}=1,\   \sum_{i=1}^M w_{i j} p_i=r_j,  \  \forall i, j,\vspace{1ex} \label{eq0_0_b}\\
    &\sum_{i=1}^M \sum_{j=1}^N w_{i j} p_i d_{i j} \leq D,\   \sum_{j=1}^N r_j=1 ,\vspace{1ex} \label{eq0_0_c}\\
    &d(\bm{p},\bm{r}) \leq P. \label{eq0_0_d}
\end{align}
\end{subequations}
%

\subsection{Perception Measures}
One commonly used measure of perceptual quality $d(\bm{p},\bm{r})$ is the Wasserstein metric,\par
%
\begin{small}
\begin{subequations} \label{metric}
\begin{align}
\mathcal{W} (\bm{p},\bm{r}) =  &\min_{\bm{\Pi}}\quad\sum_{i=1}^M \sum_{j=1}^N \Pi_{i j} c_{i j} \vspace{1ex}\\
      &\text { s.t. } \sum_{i=1}^M \Pi_{i j}= r_j ,\  \sum_{j=1}^N \Pi_{i j}= p_i,\  \forall i, j,
\end{align}
\end{subequations}%
\end{small}%
where $c_{ij}$ denotes the cost matrix between $x_i$ and $\hat{x}_j$. 
In some cases, the total variation (TV) distance $\delta(\bm{p}, \bm{r}) = \frac{1}{2}\|\bm{p}-\bm{r} \Vert_1$ and the Kulback-Leibler (KL) divergence $\text{KL}(\bm{p}\| \bm{r}) = \sum_{i=1}^M p_i \left[\log p_{i}-\log r_i\right]$ are also considered as a measure of perceptual quality \cite{blau2019rethinking}. We will focus on the Wasserstein metric, as generalizations to the TV and KL metrics follow straightforwardly as will be discussed in Section III.

\subsection{Wasserstein Barycenter Model for RDP functions}


%
%
%

Obviously, solving the RDP functions is more difficult than solving the RD function due to the introduction of the new perception constraint \eqref{eq0_0_d}. Numerically, we need to frequently update the $d(\bm{p},\bm{r})$ to ensure that the perception constraint is satisfied. For metrics such as the Wasserstein metric \eqref{metric}, the computational cost for this is high. Moreover, the original feasible domain of RD functions is changed due to the introduction of \eqref{eq0_0_d}. The feasible solution of RD functions is inside the simplex structure since \eqref{eq0_0_b} and \eqref{eq0_0_c} are both linear constraints. This is the reason why the BA algorithm and the AS algorithm can solve the RD functions at low costs \cite{wu2022communication}. However, the perception constraint \eqref{eq0_0_d} breaks the desirable simplex structure, which brings computational challenges. In this section, we develop a new model and an algorithm for solving RDP functions efficiently and accurately. Our starting point is to convert the perception constraint \eqref{eq0_0_d} into a linear constraint in a higher dimensional space. Consequently, all the constraints, \eqref{eq0_0_b}-\eqref{eq0_0_d}, also preserve the simplex structure in the higher-dimensional space. This would bring great convenience to our algorithm design. The main idea is summarized into the following theorem:

\begin{theorem}
\label{thm-0}
The solution to RDP functions, which is the optimal value of \eqref{eq0_0}, equals the optimal value of the following optimization problem. Meanwhile, the optimal solution $(\bm{w},\bm{r})$ of the following optimization problem \eqref{eq0_1} is the optimal solution of \eqref{eq0_0}:\par
 \begin{small}
 \begin{subequations} \label{eq0_1}
\begin{align}
    \min _{\bm{w},\bm{r},\bm{\Pi}} \quad  &\sum_{i=1}^M \sum_{j=1}^N\left(w_{i j} p_i\right)\left[\log w_{i j}-\log r_j\right] \vspace{1ex}\label{eq0_1_a}\\
    \emph{ \text { s.t. } } &\sum_{j=1}^N w_{i j}=1,\   \sum_{i=1}^M w_{i j} p_i=r_j, \vspace{1ex}\label{eq0_1_b}\\
    &\sum_{i=1}^M \Pi_{i j}= r_j ,\  \sum_{j=1}^N \Pi_{i j}= p_i, \  \forall i, j,\vspace{1ex}\label{eq0_1_c}\\
    &\sum_{i=1}^M \sum_{j=1}^N w_{i j} p_i d_{i j} \leq D,\   \sum_{j=1}^N r_j=1 ,\vspace{1ex}\label{eq0_1_d}\\
    &\sum_{i=1}^M \sum_{j=1}^N \Pi_{i j} c_{i j} \leq P.\label{eq0_1_e}
\end{align}
\end{subequations}
 \end{small}
\end{theorem}

    
    
%

Theorem \ref{thm-0} establishes an equivalence between the RDP problem \eqref{eq0_0} and the optimization \eqref{eq0_1}. Also, we observe that the model \eqref{eq0_1} has the Wasserstein Barycenter structure. The optimization variable $r_j$ can be regarded as the Barycenter, and $\text{diag}(\bm{p})\cdot\bm{w}$ and $\bm{\Pi}$ are two couplings between each input $\bm{p}$ and Barycenter $\bm{r}$. Thus, we have obtained the Wasserstein Barycenter model for RDP functions. In general, the numerical methods for the Wasserstein Barycenter problem are computationally intensive \cite{DBLP:conf/icml/CuturiD14, peyre2019computational}. However, as $\bm{\Pi}$ is not explicitly included in the objective function of \eqref{eq0_1}, it is possible to design a fast and efficient algorithm for the above WBM-RDP.

\section{Entropy Regularized WBM-RDP and Improved AS Algorithm}

The above section establishes the WBM-RDP model \eqref{eq0_1}, which is the first step towards computing the RDP functions. However, there are two difficulties in designing a practical algorithm. First, the WBM-RDP \eqref{eq0_1} is not strictly convex on $\bm{\Pi}$. Although the optimal value of \eqref{eq0_1} is unique, the corresponding optimal solutions may vary in the dimensions of $\bm{\Pi}$. Therefore, the convergence and the numerical stability of the AS algorithm designed for RD functions \cite{wu2022communication} cannot be guaranteed. Second, WBM-RDP contains logarithmic terms of the Barycenter objective optimization function $\log r_j$, which is not standard formulation of the classical Wasserstein Barycenter problems. In this section, we will overcome these difficulties and improve the Alternating Sinkhorn algorithm to solve WBM-RDP.
%
%

%
%

\subsection{Entropy Regularized WBM-RDP}
%
As discussed above, the WBM-RDP \eqref{eq0_1} is not strictly convex on $\bm{\Pi}$, the most direct way \cite{nutz2022entropic} is to introduce an extra entropy regularized term in the objective optimization function \eqref{eq0_1_a}, i.e.,\par
\begin{small}
\begin{equation*}
H(\bm{\Pi})= \sum_{i=1}^M \sum_{j=1}^N \Pi_{i j} \log(\Pi_{i j}).
\end{equation*}%
\end{small}%
%
This leads to the entropy regularized WBM-RDP:\par
%
\begin{subequations}\label{problem}
\begin{small}
\begin{align}
    \min _{\bm{w},\bm{r},\bm{\Pi}} \quad &\sum_{i=1}^M \sum_{j=1}^N\left(w_{i j} p_i\right)\left[\log w_{i j}-\log r_j\right]  
    + \varepsilon H(\bm{\Pi})   \vspace{1ex}\\
    \text { s.t. }  &\sum_{j=1}^N w_{i j}=1,\   \sum_{i=1}^M w_{i j} p_i=r_j, \vspace{1ex}\\
    &\sum_{i=1}^M \Pi_{i j}= r_j ,\  \sum_{j=1}^N \Pi_{i j}= p_i,  \forall i, j. \label{problem_c}\vspace{1ex}\\
    &\sum_{i=1}^M \sum_{j=1}^N w_{i j} p_i d_{i j} \leq D,\   \sum_{j=1}^N r_j=1 ,\label{problem_d}\vspace{1ex}\\
    &\sum_{i=1}^M \sum_{j=1}^N \Pi_{i j} c_{i j} \label{problem_e}\leq P.
\end{align}
\end{small}
\end{subequations}
Here, $\varepsilon>0$ is a newly introduced regularization parameter. Thus, we get the entropy regularized WBM-RDP with strict convexity. Moreover, during the alternative optimization iterations, closed-form expressions of the dual variables can always be obtained, which accelerates the algorithm while improving the accuracy. Not only that, the following theorem guarantees that the solution to entropy regularized WBM-RDP \eqref{problem} converges to WBM-RDP \eqref{eq0_1} as $\varepsilon \rightarrow 0$.
\begin{theorem}
\label{thm-1}
\emph{(Convergence in $\varepsilon$)} The solution $\{ \bm{w}_{\varepsilon},\bm{\Pi}_{\varepsilon},\bm{r}_{\varepsilon} \}$ to (\ref{problem}) converges to the optimal solution with minimal entropy of $H(\bm{\Pi})$ within the set of all optimal solutions to (\ref{eq0_1}), i.e, 
\begin{equation}\label{proof1}
   \{ \bm{w}_{\varepsilon},\bm{\Pi}_{\varepsilon},\bm{r}_{\varepsilon} \} \underset{\varepsilon \rightarrow 0}{\longrightarrow} \operatorname{argmin} \Big \{H(\bm{\Pi}) \Big| \left\{ \bm{w},\bm{\Pi},\bm{r} \right\} \in \mathcal{M}\Big \},
\end{equation}
where $\mathcal{M}$ denotes the set of all optimal solutions to (\ref{eq0_1}).
\end{theorem}

\subsection{The Improved Alternating Sinkhorn Algorithm}

 We construct the Lagrangian function of the regularized WBM-RDP \eqref{problem} by introducing dual variables $\boldsymbol{\alpha},\boldsymbol{\theta} \in \mathbb{R}^M, \boldsymbol{\beta},\boldsymbol{\tau} \in \mathbb{R}^N, \lambda,\gamma \in \mathbb{R}^{+}$ and $\eta \in \mathbb{R}$:\par
 \begin{small}
\begin{equation}\label{Lagrangian}
\begin{aligned}
    &\begin{aligned}
    &\mathcal{L} \left(\bm{w},\bm{\Pi},\bm{r}; \bm{\alpha},\bm{\beta}, \bm{\theta}, \bm{\tau}, \lambda,\eta, \gamma\right) \\
    &=\sum_{i=1}^M{\sum_{j=1}^N} w_{i j} p_i \log \frac{w_{i j}}{r_j}+\varepsilon \sum_{i=1}^M \sum_{j=1}^N \Pi_{i j} \ln \Pi_{i j}
    \end{aligned}\\
    &+\sum_{i=1}^M \alpha_i\left(\sum_{j=1}^N w_{i j}-1\right)+\sum_{j=1}^N \beta_j\left(\sum_{i=1}^N w_{i j} p_i-r_j\right)\\
    &+\sum_{i=1}^M \theta_i\left(\sum_{j=1}^N \Pi_{i j}-p_i\right)+\sum_{j=1}^N \tau_j\left(\sum_{i=1}^M \Pi_{i j}-r_j\right)\\
    &+\lambda\left(\sum_{i=1}^M \sum_{j=1}^N w_{i j} p_i d_{i j}-D\right)+\eta\left(\sum_{j=1}^N r_j-1\right)\\
    &+\gamma\left(\sum_{i=1}^M \sum_{j=1}^N \Pi_{i j} c_{i j}-P\right).
\end{aligned}
\end{equation}
\end{small}

Here we note that the Lagrangian function \eqref{Lagrangian} is convex with respect to each variable. 
Furthermore, we can improve the algorithm by designing the directions of the alternating iterations according to how the variables appear in \eqref{Lagrangian}.
Next, we sketch the main ingredients of our algorithm. 
The pseudo-code is presented in Algorithm \ref{alg:OT_rdp}.

%

\begin{enumerate}
  \item [1)] 
   Update $\bm{w}$ and associated dual variables $\boldsymbol{\alpha,\beta},\lambda$ while fixing  $\bm{r}, \bm{\Pi}$ as constant parameters. First, we can update $\alpha_i$ and $\beta_j$:\par
   \begin{small}
   \begin{equation}
    \psi_j \leftarrow 1 \Big/ \sum_{i=1}^M K_{i j} \phi_i p_i, \quad \phi_i \leftarrow 1 \Big/ \sum_{j=1}^N K_{i j} \psi_j r_j,
\end{equation}
\end{small}
where $\phi_i=\exp (-\alpha_i / p_i-1/ 2), \psi_j = \exp (-\beta_j-1/2)$ and  $K_{i j}=\exp (-\lambda d_{i j})$. Second, we update $\lambda$ by a root-searching operation on the following monotonic single-variable function on $\mathbb{R}^{+}$:\par
\begin{small}
\begin{equation}
    F(\lambda) \triangleq \sum_{i=1}^M\sum_{j=1}^N d_{i j}p_i \phi_i e^{-\lambda d_{i j}} \psi_j r_j - D = 0.
\end{equation}
\end{small}
  \item [2)]
   Update $\bm{\Pi}$ and associated dual variables $\boldsymbol{\theta,\tau},\gamma$ while fixing  $\bm{r}, \bm{w}$ as constant parameters. We alternatively update $\theta_i, \tau_j$:\par
   \begin{small}
   \begin{equation}
    \varphi_j \leftarrow r_j \Big/ \sum_{i=1}^M M_{i j} \xi_i, \quad \xi_i \leftarrow p_i \Big/ \sum_{j=1}^N M_{i j} \varphi_j,
\end{equation}
\end{small}
where $\xi _i=\exp(-\theta_i / \varepsilon-1/2), \varphi _j = \exp(-\tau_j/ \varepsilon-1/2)$ and $M_{i j}=\exp(-\gamma  c_{i j}/ \varepsilon)$. Again, we can update $\gamma$ by the root-searching operation on the following monotonic single-variable function on $\mathbb{R}^{+}$:\par
\begin{small}
\begin{equation}
    G(\gamma) \triangleq  \sum_{i=1}^M\sum_{j=1}^N c_{i j} \xi_i e^{-\gamma  c_{i j}/ \varepsilon} \varphi_j - P = 0 .
\end{equation}%
\end{small}%
  \item [3)]
   Update $\bm{r}$ and associated dual variables $\eta$ while fixing $\bm{w}, \bm{\Pi}$ as constant parameters. We can update $\eta$ by finding the root of the following single-variable function on its largest monotone interval $(\max_j(\beta_j+\tau_j),+\infty)$:\par
   \begin{small}
   \begin{equation}
    H(\eta) \triangleq \sum_{j=1}^N\left[\left(\sum_{i=1}^M w_{i j} p_i\right) \Big/\left(\eta-\beta_j-\tau_j\right)\right]-1=0,
\end{equation}
\end{small}
and we finally get 
   \begin{small}
   \begin{equation}
 r_j =\left(\sum_{i=1}^M w_{i j} p_i\right) \Big/ (\eta-\beta_j-\tau_j). 
  \end{equation}
\end{small}
 \end{enumerate}


We stress that the three steps above do not contain inner iterations, because $\bm{\Pi}$ is not explicitly included in the objective optimization function of the WBM-RDP as discussed above. 
Therefore, compared to other existing algorithms for solving the Wasserstein Barycenter, our proposed algorithm gain much efficiency and simplicity. 
On the other hand, our proposed algorithm adopts a similar alternating iteration technique to the Alternating Sinkhorn algorithm for RD functions. 
However, a vital operation we conduct is altering the projection direction of different joint distribution variables, i.e., $\bm{w}$ and $\bm{\Pi}$, which is essentially different from the AS algorithm for RD functions. 
Thus we name it the Improved Alternating Sinkhorn Algorithm. 
%

\begin{algorithm}[t]
\begin{small}
 \caption{The Improved Alternating Sinkhorn Algorithm}
 \label{alg:OT_rdp}
 \begin{algorithmic}[1]
  \REQUIRE Distortion measure $d_{ij}$, marginal distribution $p_{i}$,\\cost matrix $c_{ij}$, regularization parameter $\varepsilon$, \\maximum iteration number $max\_iter$.
  %
  \ENSURE Minimal value  $\sum_{i=1}^{M} \sum_{j=1}^{N} (w_{i j}p_{i}) \left[\log w_{i j}-\log r_{j}\right]+ \varepsilon \sum_{i=1}^M \sum_{j=1}^N \Pi_{i j} \log(\Pi_{i j})$  with respect to variables $\bm{w}$, $\bm{r}$ and $\bm{\Pi}$.
  \STATE \textbf{Initialization:} $\bm{\phi}, \bm{\xi} = \mathbf{1}_{M}, \bm{\psi},\bm{\varphi} = \mathbf{1}_{N}, \lambda,\gamma=1;$
  \STATE Set $K_{ij} \gets \exp(-\lambda d_{ij})$
  \STATE Set $M_{ij} \gets \exp(-\gamma c_{ij}/\varepsilon)$
  \FOR{$\ell = 1 : max\_iter$}
  \STATE $\psi_{j} \gets 1/\sum_{i=1}^{M}K_{ij}\phi_{i}p_{i}, \quad j=1,\cdots,N$
  \STATE $\phi_{i} \gets 1/\sum_{j=1}^{N}K_{ij}\psi_{j}r_{j}, \quad i=1,\cdots,M$
  \STATE Solve $F(\lambda) = 0$ for $\lambda\in\mathbb{R}^{+}$ with Newton's method
  \STATE Update $K_{ij} \gets \exp(-\lambda d_{ij})$ and $w_{ij}\gets\phi_{i}K_{ij}\psi_{j}r_{j}$
  \STATE $\varphi_j \gets r_j / \sum_{i=1}^M M_{i j} \xi_i, \quad j=1,\cdots,N$
  \STATE $\xi_i \gets p_i / \sum_{j=1}^N M_{i j} \varphi_j, \quad i=1,\cdots,M$
  \STATE Solve $G(\gamma) = 0$ for $\gamma\in\mathbb{R}^{+}$ with Newton's method
  \STATE Update $M_{ij} \gets \exp(-\gamma c_{ij}/\varepsilon)$ 
  \STATE Solve $H(\eta) = 0$ for $\eta\in\mathbb{R}$ with Newton's method
  \STATE Update $r_{j}\gets\left(\sum_{i=1}^{M} w_{i j}p_{i} \right)\Big/\left(\eta-\beta_{j}-\tau_{j}\right)$ 
  \ENDFOR
  \STATE \textbf{end}
  \RETURN $\sum_{i=1}^{M}\sum_{j=1}^{N} \left(\phi_{i}p_{i}K_{ij}\psi_{j}r_{j}\right)\left[\log \left(\phi_{i}K_{ij}\psi_{j}\right)\right]$
 \end{algorithmic}
 \end{small}
\end{algorithm}



\begin{remark}
    If the perception measure in \eqref{eq0_0_d} is substituted by TV distance, we only need to set the cost matrix in \eqref{eq0_1_e} as $c_{ij} = \mathbf{1}_{i \neq j}$ (see Eq. (6.11) of \cite{villani2009optimal}). Then our improved AS algorithm is still applicable.   
\end{remark}

\begin{remark}
   If the perception measure in \eqref{eq0_0_d} is substituted by KL divergence, our improved AS algorithm would be simpler. The Sinkhorn iteration in step 2) can be omitted since $\bm{\Pi}$ does not need to be introduced. In step 2) the $G(\gamma)$ is substituted by \par 
    \begin{small}
    \begin{equation*}
       G(\gamma) \triangleq \sum_{j=1}^M p_j \log \left(\Big(\gamma p_j + \sum_{i=1}^M w_{i j} p_i \Big) \Big/(\eta -\beta_j)\right) - T,
    \end{equation*}%
    \end{small}%
    where $T = \sum_{i=1}^M p_i \log p_{i} - P$ and $H(\eta)$ is substituted by\par
    \begin{small}
    \begin{equation*}
        H(\eta) \triangleq \sum_{j=1}^M \left(\gamma p_j + \sum_{i=1}^M w_{i j} p_i \right) \Big/(\eta -\beta_j)-1,
    \end{equation*}%
    \end{small}%
     and $r_j = (\gamma p_j + \sum_{i=1}^M w_{i j} p_i ) /(\eta -\beta_j) $. 

   
\end{remark}



\section{Numerical Experiment}


In this section, we numerically study the validity of the WBM-RDP and the improved AS algorithm. All the experiments are conducted on a PC with 8G RAM, and one 11th Gen Intel (R) Core (TM) i5-1135G7 CPU @2.40GHz.

We compute RDP functions under two settings with different perception measures: one is the binary source with Hamming distortion \cite{blau2019rethinking}, and the other is the Gaussian source with squared error distortion \cite{zhang2021universal}. Moreover, the above two settings have analytical expressions \cite{blau2019rethinking,zhang2021universal} when the perceptual constraints are TV distance and Wasserstein-2 metric, respectively.


%
%

For the binary source, we can directly compute the result since we can set the discrete distribution $p$ beforehand. As for Gaussian source, we first truncate the sources into an interval $[-S, S]$ and then discretize it by a set of uniform grid points $\{ x_i\}_{i=1}^{N}$ whose adjacent spacing is $\delta = 2S/({N-1})$, i.e.,
\begin{equation*}
    x_i = -S + (i-1)\delta, \quad i = 1,\cdots, N.
\end{equation*}
The corresponding distribution $\bm{p}$ of the Gaussian source can then be denoted by
\begin{equation}
p_i = F(x_i+\frac{\delta}{2})-F(x_i-\frac{\delta}{2}), \quad i = 1,\cdots, N,
\end{equation}
where $F(x)$ denotes the distribution of the Gaussian source. Unless otherwise specified, we take $p = 0.1$ for the binary source and $S = 8, \delta = 0.5, \mu = 0, \sigma = 2$ for the Gaussian source. For our method, we set $\varepsilon = 0.01$. 
Additionally, we note that the space complexity is $\mathcal{O}(N^2)$ with the dimension of the above discretized Gaussian source. 
The following results will suggest $\delta = 0.5$ is precise enough for such discretization. 

\begin{figure}[t]
\centering
\includegraphics[width=0.8\linewidth]{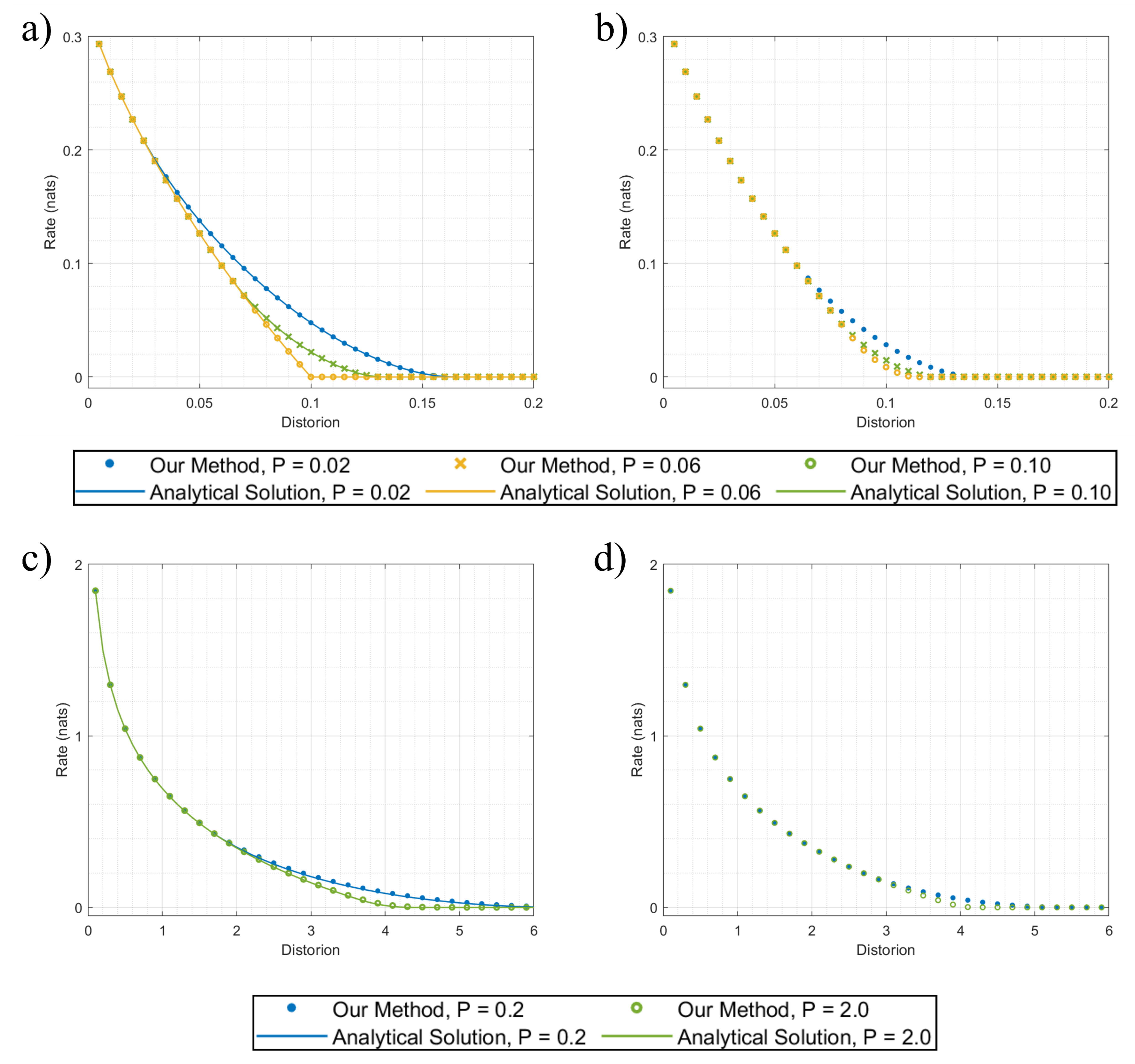}
\captionsetup{font=scriptsize}
\caption{The Rate-Distortion-Perception functions obtained by our method and a) the binary source with TV distance and its analytical solution; b) the binary source with KL divergence; c) the Gaussian source with Wasserstein-2 metric and its analytical solution;  d) the Gaussian source with TV distance.}  \label{fig1}
\end{figure}

\begin{figure}[t]
\centering
\includegraphics[width=0.8\linewidth]{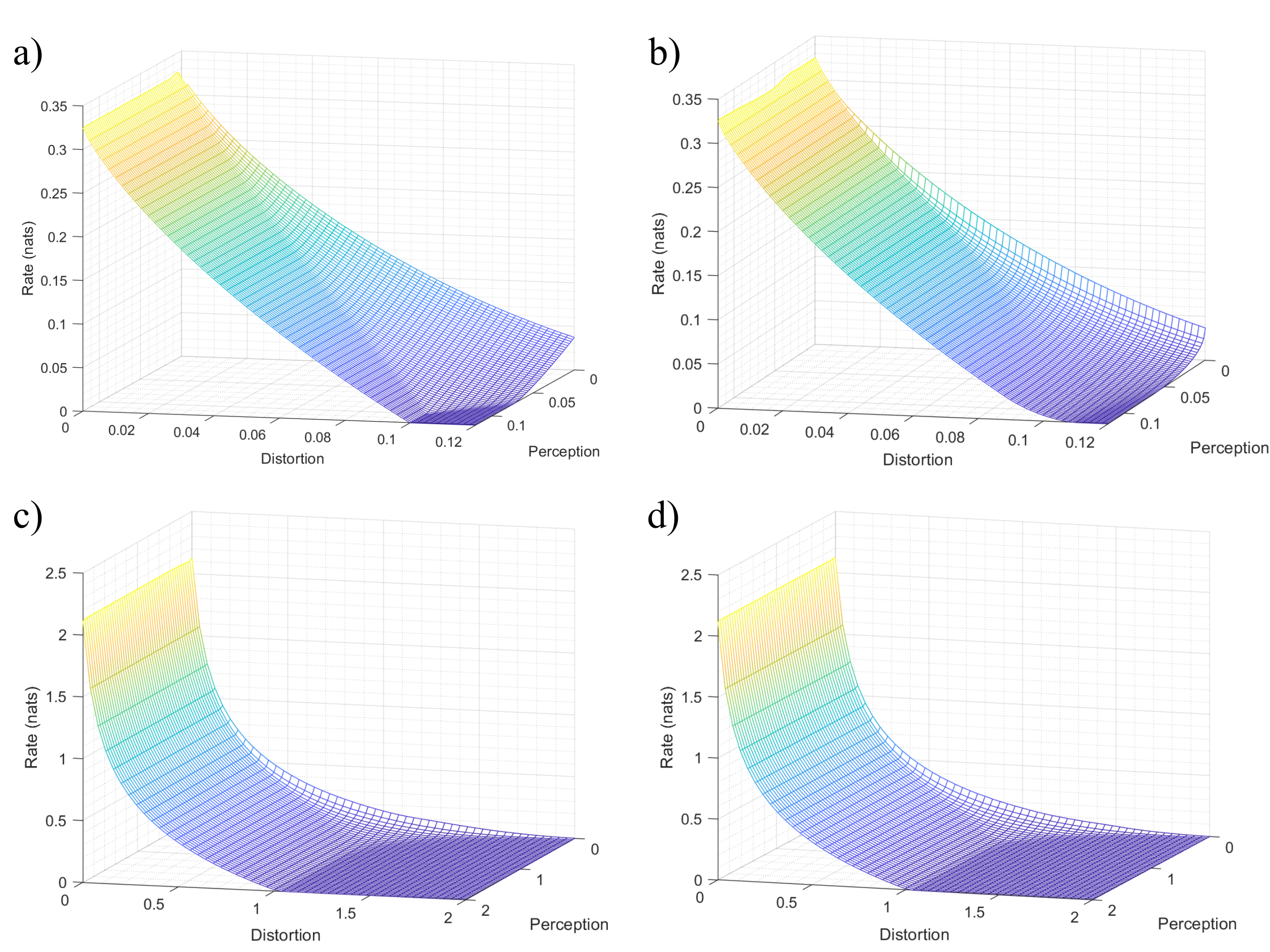}
\captionsetup{font=scriptsize}
\caption{The 3D plot of the Rate-Distortion-Perception functions obtained by
our method. a) the
binary source with TV distance. b) the binary source with KL
divergence. c) the Gaussian source with Wasserstein-2 metric d) the Gaussian source with TV distance.}  \label{fig2}
\end{figure}

In Fig. \ref{fig1}, we plot RDP curves given by our method under different perception parameter $P$ and compare them to the results with known theoretical expression. The results obtained by our method match well the analytic expression in Fig. \ref{fig1} a) and c). Furthermore, we can also plot the results where the analytical solution is not known in Fig. \ref{fig1} b) and d). 
We also plot the 3D diagram of RDP surface in Fig. \ref{fig2}. For the Gaussian source, we set $S = 4, \sigma = 1$ for visual effect. The results are in accord with those derived from data-driven methods in \cite{blau2019rethinking}.

\subsection{Algorithm Convergence Verification}

We verify the convergence of the improved AS algorithm in this subsection. Here we consider the residual errors of the Karish-Kuhn-Tucker (KKT) condition of the optimization problem (\ref{problem}) to be the indicator of convergence. We define $L_1$ residual errors $r_{\psi}$ as $ r_{\psi}=\sum_{j=1}^N\left|\psi_j \sum_{i=1}^M K_{i j} \phi_i p_i- 1\right|,$ 
and $r_{\phi}$,$r_{\lambda}$,$r_{\eta}$,$r_{\varphi}$,$r_{\xi}$,$r_{\gamma}$ can be defined similarly. We define the overall residual error $r$ to be the root mean square of the above residual errors.

\begin{figure}[t]
\centering
\includegraphics[width=0.8\linewidth]{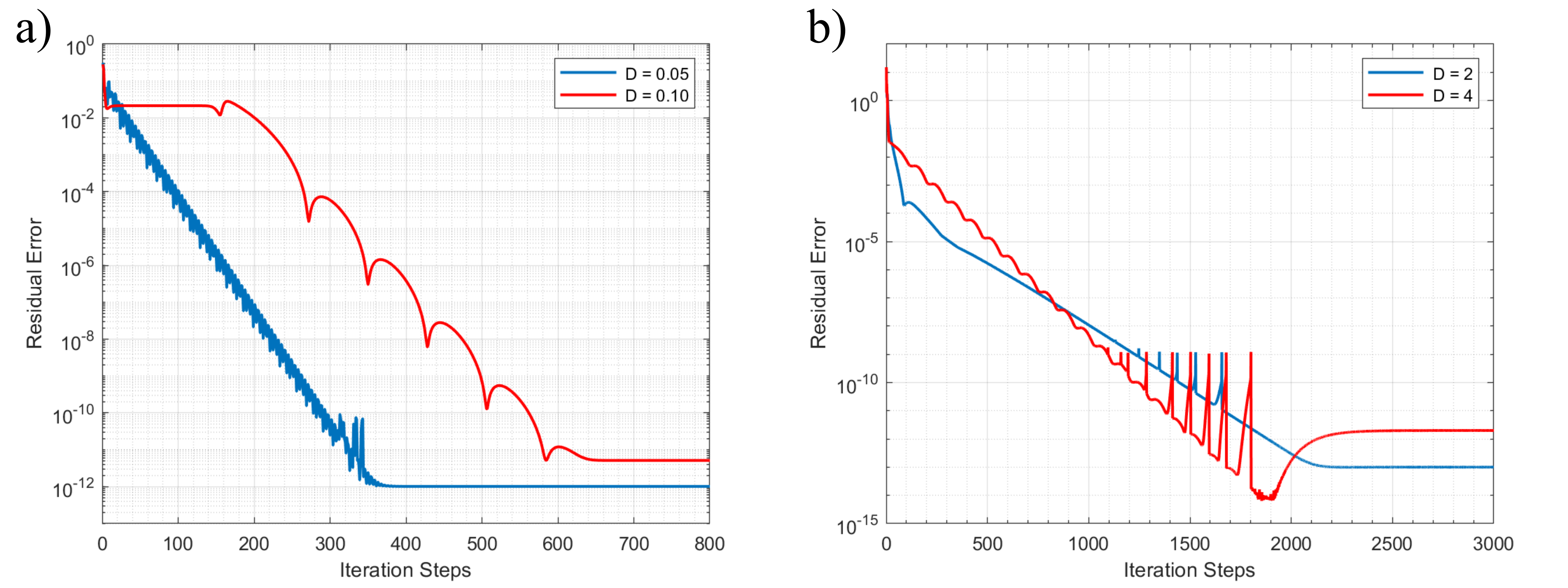}
\captionsetup{font=scriptsize}
\caption{The convergent trajectories of $r$ from our method and a) the binary source with different distortion parameters; b) the Gaussian source with different distortion parameters.}  \label{fig3}
\end{figure}

In Fig. \ref{fig3}, we respectively output the convergent trajectories of $r$ of the binary source with TV distance and Gaussian source with Wasserstein-2 metric against iteration numbers. For the binary source, we set $P = 0.06$. For the Gaussian source, we set $ P = 2$. The results show different convergence behaviors with different distortion parameters, and all of these cases will converge below $1 \times 10^{-10}$ at last. 

\subsection{ The Regularization Parameter $\varepsilon$ }


We have theoretically demonstrated that when $\varepsilon \rightarrow 0$ the entropy regularized problem converges to the original problem. However, according to the general results of entropy regularized OT problem, problems with smaller $\varepsilon$ have higher precision but require more computation \cite{peyre2019computational}. 
Moreover, according to the theory on Sinkhorn \cite{peyre2019computational}, an excessive $\lambda$ might trigger numerical stability problems.
Thus we wish to investigate the impact of $\varepsilon$ on the WBM-RDP.

Results are shown in Table \ref{Tabel2}, where the error represents the $L_1$ difference between the algorithm results and the explicit results. Here for the binary source with TV distance, we set $P = 0.06$. And we set $P = 2$ for the Gaussian source with Wasserstein-2 metric. 
When $\varepsilon$ decreases, the computational time rises while the results are more accurate regardless of the source. 
Furthermore, for $\varepsilon = 1e-4$ in both sources, the small $\varepsilon$ causes numerical instability. 
Therefore, from a practical perspective, $\varepsilon = 0.01$ seems to be an ideal choice for RDP functions, as it ensures a certain level of accuracy and does not consume too much time.

\begin{table}[H]
\begin{tiny}
\centering
\captionsetup{font=scriptsize}
\caption{The computational time and error against $\varepsilon$ with different sources. }
\label{Tabel2}
\begin{center}
\resizebox{0.9\columnwidth}{!}{
\begin{tabular}{c|l|c|c}
\hline
\multicolumn{1}{l|}{}            & \multicolumn{1}{c|}{$\varepsilon$} & Time (s) & Error \\ \hline
\multirow{5}{*}{Binary source}   & 1.00E-01               & 6.16                 & 4.36E-04           \\
                                 & 5.00E-02               & 6.50                 & 8.75E-05           \\
                                 & 1.00E-02               & 8.69                 & 5.41E-06           \\
                                 & 5.00E-03               & 10.62                & 3.47E-06           \\ 
                                 & 1.00E-04                & -
                                   & -                  \\ \hline
\multirow{5}{*}{Gaussian source} & 1.00E-01               & 70.01                & 1.49E-02           \\
                                 & 5.00E-02               & 74.87                & 7.30E-03           \\
                                 & 1.00E-02               & 99.56                & 2.30E-03           \\
                                 & 5.00E-03               & 118.97               & 1.80E-03           \\ 
                                 & 1.00E-04                & -
                                   & -                  \\ \hline
\end{tabular}}
\end{center}
\end{tiny}
\end{table}

\section{Conclusion}
In this work, we present a new scheme for computing the information Rate-Distortion-Perception functions. We convert the original problem to a Wasserstein Barycenter model for Rate-Distortion-Perception functions. Furthermore, we propose the improved Alternating Sinkhorn method with entropy regularization to solve the optimization problem. Numerical experiments show that our algorithm performs with high accuracy and efficiency. 
%
Extensions to properties of RDP functions and implementation to practical lossy compression schemes will be reported in the journal version.


\IEEEtriggeratref{11}
\bibliography{ref.bib} 

\begin{thebibliography}{10}
\providecommand{\url}[1]{#1}
\csname url@samestyle\endcsname
\providecommand{\newblock}{\relax}
\providecommand{\bibinfo}[2]{#2}
\providecommand{\BIBentrySTDinterwordspacing}{\spaceskip=0pt\relax}
\providecommand{\BIBentryALTinterwordstretchfactor}{4}
\providecommand{\BIBentryALTinterwordspacing}{\spaceskip=\fontdimen2\font plus
\BIBentryALTinterwordstretchfactor\fontdimen3\font minus
  \fontdimen4\font\relax}
\providecommand{\BIBforeignlanguage}[2]{{%
\expandafter\ifx\csname l@#1\endcsname\relax
\typeout{** WARNING: IEEEtran.bst: No hyphenation pattern has been}%
\typeout{** loaded for the language `#1'. Using the pattern for}%
\typeout{** the default language instead.}%
\else
\language=\csname l@#1\endcsname
\fi
#2}}
\providecommand{\BIBdecl}{\relax}
\BIBdecl

\bibitem{DBLP:conf/nips/MentzerTTA20}
F.~Mentzer, G.~Toderici, M.~Tschannen, and E.~Agustsson, ``High-fidelity
  generative image compression,'' in \emph{Advances in Neural Information
  Processing Systems (NeurIPS 2020)}, H.~Larochelle, M.~Ranzato, R.~Hadsell,
  M.~Balcan, and H.~Lin, Eds., vol.~33, Dec. 2020, pp. 11\,913--11\,924.

\bibitem{DBLP:journals/tcsv/MaZJZWW20}
S.~Ma, X.~Zhang, C.~Jia, Z.~Zhao, S.~Wang, and S.~Wang, ``Image and video
  compression with neural networks: {A} review,'' \emph{IEEE Transactions on
  Circuits and Systems for Video Technology}, vol.~30, no.~6, pp. 1683--1698,
  2020.

\bibitem{lu2020end}
G.~Lu, X.~Zhang, W.~Ouyang, L.~Chen, Z.~Gao, and D.~Xu, ``An end-to-end
  learning framework for video compression,'' \emph{IEEE Transactions on
  Pattern Analysis and Machine Intelligence}, vol.~43, no.~10, pp. 3292--3308,
  2020.

\bibitem{DBLP:journals/taslp/ZeghidourLOST22}
N.~Zeghidour, A.~Luebs, A.~Omran, J.~Skoglund, and M.~Tagliasacchi,
  ``Soundstream: An end-to-end neural audio codec,'' \emph{IEEE/ACM
  Transactions on Audio, Speech, and Language Processing}, vol.~30, pp.
  495--507, 2022.

\bibitem{shannon1959coding}
C.~E. Shannon \emph{et~al.}, ``{Coding Theorems for a Discrete Source with a
  Fidelity Criterion},'' \emph{Institute of Radio Engineers International
  Convention Record}, vol.~4, no. 142-163, p.~1, Mar. 1959.

\bibitem{DBLP:books/wi/01/CT2001}
T.~M. Cover and J.~A. Thomas, \emph{{Elements of Information Theory}}.\hskip
  1em plus 0.5em minus 0.4em\relax Wiley-Interscience, 2006.

\bibitem{DBLP:journals/corr/SanturkarBS17}
S.~Santurkar, D.~M. Budden, and N.~Shavit, ``{Generative Compression},'' in
  \emph{2018 Picture Coding Symposium (PCS)}, San Francisco, California, USA,
  Jun. 2018, pp. 1--5.

\bibitem{DBLP:conf/iccv/AgustssonTMTG19}
E.~Agustsson, M.~Tschannen, F.~Mentzer, R.~Timofte, and L.~V. Gool,
  ``{Generative Adversarial Networks for Extreme Learned Image Compression},''
  in \emph{2019 {IEEE/CVF} International Conference on Computer Vision (ICCV)},
  Seoul, South Korea, Oct.-Nov. 2019, pp. 221--231.

\bibitem{DBLP:conf/cvpr/BlauM18}
Y.~Blau and T.~Michaeli, ``{The Perception-Distortion Tradeoff},'' in
  \emph{2018 IEEE/CVF Conference on Computer Vision and Pattern Recognition
  (CVPR)}, Salt Lake City, Utah, USA, Jun. 2018, pp. 6228--6237.

\bibitem{blau2019rethinking}
\vspace{0mm}Y. Blau and T.~Michaeli, ``{Rethinking Lossy Compression: The
  Rate-Distortion-Perception Tradeoff},'' in \emph{36th International
  Conference on Machine Learning (ICML)}, Long Beach, California, USA, Jun.
  2019, pp. 675--685.

\bibitem{zhang2021universal}
G.~Zhang, J.~Qian, J.~Chen, and A.~Khisti, ``{Universal
  Rate-Distortion-Perception Representations for Lossy Compression},'' in
  \emph{Advances in Neural Information Processing Systems (NIPS 2021)},
  vol.~34, Dec. 2021, pp. 11\,517--11\,529.

\bibitem{niu2023conditional}
X.~Niu, D.~Gündüz, B.~Bai, and W.~Han, ``Conditional
  rate-distortion-perception trade-off,'' in \emph{2023 IEEE International
  Symposium on Information Theory (ISIT)}.\hskip 1em plus 0.5em minus
  0.4em\relax (to appear), 2023.

\bibitem{DBLP:journals/tit/Arimoto72}
S.~Arimoto, ``{An Algorithm for Computing the Capacity of Arbitrary Discrete
  Memoryless Channels},'' \emph{IEEE Transactions on Information Theory},
  vol.~18, no.~1, pp. 14--20, Jan. 1972.

\bibitem{DBLP:journals/tit/Blahut72}
R.~E. Blahut, ``{Computation of Channel Capacity and Rate-Distortion
  Functions},'' \emph{IEEE Transactions on Information Theory}, vol.~18, no.~4,
  pp. 460--473, Jan. 1972.

\bibitem{DBLP:conf/pcs/KirmemisT21}
O.~Kirmemis and A.~M. Tekalp, ``{A Practical Approach for
  Rate-Distortion-Perception Analysis in Learned Image Compression},'' in
  \emph{2021 Picture Coding Symposium (PCS)}, Bristol, United Kingdom, Jun.
  2021, pp. 1--5.

\bibitem{wu2022communication}
S.~Wu, W.~Ye, H.~Wu, H.~Wu, W.~Zhang, and B.~Bai, ``{A Communication Optimal
  Transport Approach to the Computation of Rate Distortion Functions},''
  \emph{arXiv preprint arXiv:2212.10098}, 2022.

\bibitem{pass2015multi}
B.~Pass, ``{Multi-marginal Optimal Transport: Theory and Applications},''
  \emph{ESAIM: Mathematical Modelling and Numerical Analysis}, vol.~49, no.~6,
  pp. 1771--1790, Feb. 2015.

\bibitem{DBLP:conf/icml/CuturiD14}
M.~Cuturi and A.~Doucet, ``{Fast Computation of Wasserstein Barycenters},'' in
  \emph{31th International Conference on Machine Learning (ICML)}, vol.~32,
  Beijing, China, Jun. 2014, pp. 685--693.

\bibitem{DBLP:journals/siamma/AguehC11}
M.~Agueh and G.~Carlier, ``{Barycenters in the Wasserstein Space},''
  \emph{Society for Industrial and Applied Mathematics Journal on Mathematical
  Analysis}, vol.~43, no.~2, pp. 904--924, Jan. 2011.

\bibitem{peyre2019computational}
G.~Peyr{\'e}, M.~Cuturi \emph{et~al.}, ``{Computational Optimal Transport: With
  Applications to Data Science},'' \emph{Foundations and
  Trends{\textregistered} in Machine Learning}, vol.~11, no. 5-6, pp. 355--607,
  Feb. 2019.

\bibitem{nutz2022entropic}
M.~Nutz and J.~Wiesel, ``{Entropic Optimal Transport: Convergence of
  Potentials},'' \emph{Probability Theory and Related Fields}, vol. 184, no.~1,
  pp. 401--424, Nov. 2022.

\bibitem{villani2009optimal}
C.~Villani, \emph{{Optimal Transport: Old and New}}.\hskip 1em plus 0.5em minus
  0.4em\relax Springer, 2009, vol. 338.

\end{thebibliography}

\end{document}